\documentclass[linenumbers,twocolumn]{aastex631}

\usepackage{graphicx}
\usepackage{amsmath,amsfonts,amssymb}
\usepackage{color}

\newcommand{\td}{{\rm d}}
\newcommand{\vect}[1]{\boldsymbol{#1}}

\newcommand{\be}{\begin{equation}}
\newcommand{\ee}{\end{equation}}
\newcommand{\bea}{\begin{equation} \begin{aligned}}
\newcommand{\eea}{\end{aligned} \end{equation}}
\newcommand{\Msun}{M_\odot}

\def\lsim{\mathrel{\raise.3ex\hbox{$<$\kern-.75em\lower1ex\hbox{$\sim$}}}}
\def\gsim{\mathrel{\raise.3ex\hbox{$>$\kern-.75em\lower1ex\hbox{$\sim$}}}}

\begin{document}
\nolinenumbers

\title{Probing supermassive black hole seed scenarios with gravitational wave measurements}

\author{John Ellis}
\email{john.ellis@cern.ch}
\affiliation{King’s College London, Strand, London, WC2R 2LS, United Kingdom}
\affiliation{Theoretical Physics Department, CERN, Geneva, Switzerland}
\affiliation{Keemilise ja Bioloogilise F\"u\"usika Instituut, R\"avala pst. 10, 10143 Tallinn, Estonia}

\author{Malcolm Fairbairn}
\email{malcolm.fairbairn@kcl.ac.uk}
\affiliation{King’s College London, Strand, London, WC2R 2LS, United Kingdom}

\author{Juan Urrutia}
\email{juan.urrutia@kbfi.ee}
\affiliation{Keemilise ja Bioloogilise F\"u\"usika Instituut, R\"avala pst. 10, 10143 Tallinn, Estonia}
\affiliation{Department of Cybernetics, Tallinn University of Technology, Akadeemia tee 21, 12618 Tallinn, Estonia}

\author{Ville Vaskonen}
\email{ville.vaskonen@pd.infn.it}
\affiliation{Keemilise ja Bioloogilise F\"u\"usika Instituut, R\"avala pst. 10, 10143 Tallinn, Estonia}
\affiliation{Dipartimento di Fisica e Astronomia, Universit\`a degli Studi di Padova, Via Marzolo 8, 35131 Padova, Italy}
\affiliation{Istituto Nazionale di Fisica Nucleare, Sezione di Padova, Via Marzolo 8, 35131 Padova, Italy}

\begin{abstract} \nolinenumbers
The process whereby the supermassive black holes populating the centers of galaxies have been assembled remains to be established, with the relative importance of seeds provided by collapsed Population-III stars, black holes formed in nuclear star clusters via repeated mergers, or direct collapses of protogalactic disks yet to be determined. In this paper we study the prospects for casting light on this issue by future measurements of gravitational waves emitted during the inspirals and mergers of pairs of intermediate-mass black holes, discussing in particular the roles of prospective measurements by LISA and the proposed atom interferometers AION and AEDGE. We find that, the expected number of detectable IMBH binaries is $\mathcal{O}(100)$ for LISA and AEDGE and $\mathcal{O}(10)$ for AION in low-mass seeds scenarios and goes down to $\mathcal{O}(10)$ for LISA and below one for AEDGE and AION in high-mass seed scenarios. This allows all of these observatories to probe the parameters of the seed model, in particular, if at least a fraction of the SMBHs arises from a low-mass seed population. We also show that the measurement accuracy of the binary parameters is, in general, best for AEDGE which sees very precisely the merger of the binary.
\\~~\\
KCL-PH-TH/2023-69, CERN-TH-2023-227
\end{abstract}

\section{Introduction}

Most galaxies contain supermassive black holes (SMBHs) heavier than
$10^6 \Msun$~\citep{Kormendy:2013dxa}, and the existence of black holes (BHs) with masses
between a few and $\sim 80 \Msun$ has been established by observations 
of X-ray binaries~\citep{Remillard:2006fc} and by the measurements of gravitational waves (GWs) 
with frequencies $\sim 100$\,Hz emitted
during their mergers~\citep{LIGOScientific:2018mvr,LIGOScientific:2020ibl,LIGOScientific:2021psn}. Various other
observations point to the existence of intermediate-mass black holes (IMBHs)
with masses in the range $10^4 - 10^6 \Msun$, but their mass function and 
redshift distribution is known only very poorly~\citep{Greene:2019vlv}. This lack of information
about IMBHs impedes our understanding of how SMBHs have been assembled~\citep{Reines:2022ste}. 

The main possibilities for seeding SMBH assembly include collapsed Population-III stars~\citep{Madau:2001sc,Bromm:2002hb}, BHs formed in nuclear star clusters via repeated mergers~\citep{PortegiesZwart:2004ggg,AtakanGurkan:2003hmv,Natarajan:2020avl} or direct collapses of protogalactic disks in which fragmentation is suppressed~\citep{Sesana:2004sp,Begelman:2006db,Volonteri:2007ax,Mayer:2007vk,Tanaka:2013boa,Inayoshi:2015yqa,Izquierdo-Villalba:2023ypb}. 
All of these mechanisms are capable of reproducing the properties of the observed SMBH population for suitable values of assembly parameters such as accretion rates, but can differ significantly in their predictions for the spectrum of IMBH mergers at different redshifts, see~\cite{Volonteri:2021sfo} for an overview. Signatures of these mechanisms may include either light seeds, with masses $\sim 10^2 - 10^3 M_{\odot}$ or heavy seeds, with masses $\sim 10^4 - 10^5 M_{\odot}$, at $z\lesssim 10$.  These are currently unconstrained by data, but can in principle be probed by future GW and other measurements~\citep{Sesana:2010wy,Hartwig:2016nde,Krolik:2019nnw,Mangiagli:2020rwz,Volonteri:2020wkx,Haidar:2022mko,Dong-Paez:2023qlf}.  

The purpose of this paper is to investigate what progress can be made in distinguishing between the SMBH assembly scenarios with planned future GW experiments at frequencies larger than $0.1$\,mHz in light of recent data on nHz GWs from pulsar timing array (PTA) experiments, highlighting the potential capabilities of atom interferometers. Whereas PTAs are sensitive to the early inspiralling phase of the SMBHs heavier than $10^9 M_\odot$, the future GW experiments will probe directly the infalls and mergers of binaries lighter than $10^7 M_\odot$.

NANOGrav~\citep{NANOGrav:2023gor} and other PTA experiments~\citep{EPTA:2023sfo,Zic:2023gta,Xu:2023wog} have recently
reported the observation of a stochastic background of GWs at frequencies
in the nHz range, for which the most conservative astrophysical
interpretation is that binary systems of SMBHs are emitting them with
masses $\sim 10^9 \Msun$~\citep{NANOGrav:2023hfp,NANOGrav:2023pdq,EPTA:2023xxk,Ellis:2023owy}. Naive extrapolation of binary merger models to
lower BH masses suggests that GWs from IMBH binaries may be observable at
higher frequencies between $0.1$\,mHz and $1$\,Hz, for example
by the LISA space-borne laser interferometer experiment or atom interferometer
experiments~\citep{Ellis:2023dgf}. In principle, there are two regimes where the formation channels of the SMBHs may be distinguished. Either at $z\gtrsim7$ when the seeds are assembling and scaling relations are not that strong~\citep{Treister:2013jqa,Ricarte:2017ihq}, or by observing the low mass occupation fraction of dwarf galaxies and dark matter halos at more recent times~\citep{Ricarte:2018mzn,Volonteri:2021sfo}. The extrapolation to higher frequencies of a model that can fit the NANOGrav background~\citep{Ellis:2023owy} predicts that the majority of detectable binaries will be at $z < 7$, so the focus of this study is to constrain the latter. Our work complements analogous studies that have been performed with electromagnetic observations of active galactic nuclei (AGNs)~\citep{Kelly:2012vz,Miller:2014vta, Gallo:2019fcz,Chadayammuri:2022bjj}. 

To extend these observations to the GW spectrum and establish a multi-messenger signal, we estimate how well the space-borne laser interferometer experiment LISA~\citep{2017arXiv170200786A} and the atom interferometers AION-1km~\citep{Badurina:2019hst,Badurina:2021rgt} and AEDGE~\citep{AEDGE:2019nxb,Badurina:2021rgt} could recover the occupation fraction and the weight of the different channels contributing to the formation of the SMBHs.~\footnote{See also the related discussion of the capability of the proposed TianQin experiment in~\cite{Wang:2019ryf}.} Our approach is to generate populations of the IMBH binaries, estimate how well the binary parameters could be recovered, and then construct posteriors for the parameters of the merger rate. To extend the deduced merger rate from the NANOGrav observations, we assume that the scaling relation extends to lower masses in a light-seed scenario. Although there is evidence for this assumption~\citep{Baldassare_2020}, there is no consensus among simulations and semi-analytical models~\citep{Fontanot_2015}. However, under this assumption, we predict that laser and atom interferometers will be  to place new and competitive constraints after their first years of observation, shedding light on the origin of SMBHs.

\section{Model}

We use the extended Press-Schechter formalism~\citep{Press:1973iz,Bond:1990iw,Lacey:1993iv} to estimate of the rate $R_h$ of coalescences of galactic halos of masses $M_{1,2}$. Assuming conservatively that each of the halos includes at most one BH and that the BHs merge with the halos with probability $p_{\rm BH}$~\citep{Ellis:2023owy}, the merger rate of BHs can be written as
\bea \label{MR}
    \frac{\td R_{\rm BH}(p_{\rm BH},\vect{\theta})}{\td m_1 \td m_2} \approx p_{\rm BH} \int &  \td M_1 \td M_2 \frac{\td R_h}{\td M_1 \td M_2}\\
    &\times\prod_{j=1,2} p_{\rm occ}(m_j|M_j,z,\vect{\theta}) \,, 
\eea
where $p_{\rm occ}(m_{\rm BH}|M_v,z,\vect{\theta})$ is the occupation fraction of BHs of mass $m_{\rm BH}$ in halos of mass $M_v$ at redshift $z$, and the mechanisms for SMBH formation are characterized by a set of parameters $\vect{\theta}$. We fix the value of the merging efficiency $p_{\rm BH}$ to $0.84$, which corresponds to the best fit of the SGWB measured by the NANOGrav collaboration~\citep{Ellis:2023dgf}. This fit was obtained including environmental effects on the binary evolution. Such effects are preferred by the fit but will not affect the binaries at the frequencies considered in this study.

We model the occupation fraction $p_{\rm occ}(m_{\rm BH}|M_v,z,\vect{\theta})$ by a sum over different SMBH seed channels:
\bea \label{eq:pocc}
    p_{\rm occ}&(m_{\rm BH}|M_v,z,\vect{\theta})= \\
    &\sum_j \frac{p_j(m_{\rm BH},\vect{\theta}_j)}{\sqrt{2\pi}m_{\rm BH} \sigma}\exp\!\left[-\frac{\ln(m_{\rm BH}/\bar{m})^2}{2\sigma^2}\right] \,,
\eea
where the log-normal distribution parameterizes the halo mass-BH mass relation and $p_j$ parametrizes the low-mass cut of the massive BH population. We compute the latter by using the observed halo mass-stellar mass relation $M_*(M_v)$~\citep{Girelli:2020goz}, the mean stellar mass-BH mass relation obtained from observations of inactive galaxies\footnote{Another measurement of the stellar mass-BH mass relation comes from AGN observations. As discussed in~\cite{Ellis:2023dgf}, this relation might not reflect the full low-$z$ massive BH population but only its low-mass tail.}~\citep{Kormendy:2013dxa,2015ApJ...813...82R}, $\log_{10}(\bar{m}/M_\odot) = 8.95 + 1.4 \log_{10} (M_*/10^{11}M_\odot)$, and the observed scatter of $\sigma = 1.1$ in the latter.\footnote{Notice that we use the log-normal distribution in~\eqref{eq:pocc}, so this scatter corresponds to $0.47$\,dex as in~\cite{2015ApJ...813...82R}.} 

\begin{figure*}
    \centering
    \includegraphics[height=0.34\textwidth]{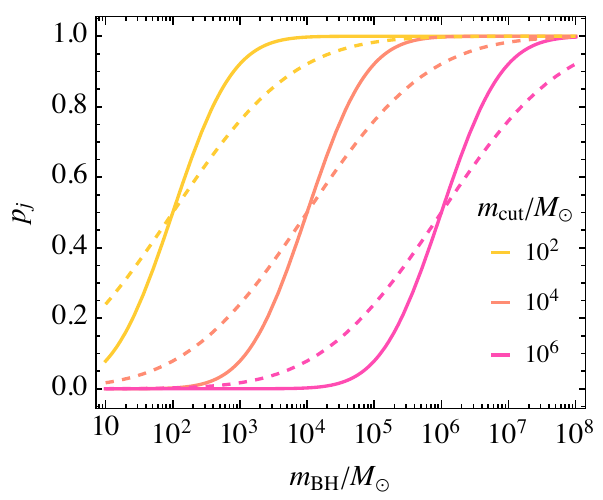} \hspace{4mm}
    \includegraphics[height=0.34\textwidth]{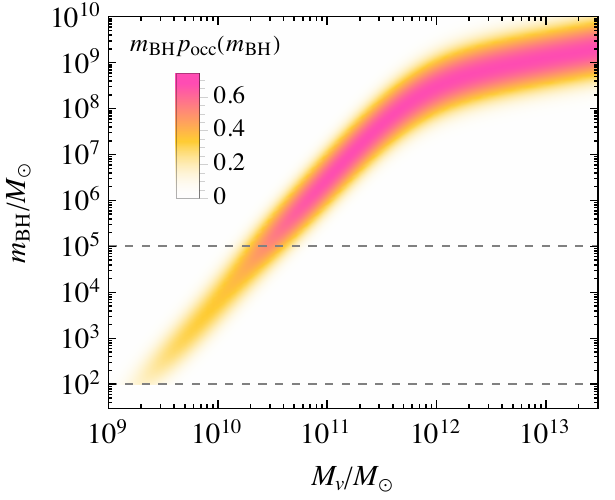}
    \caption{\emph{Left panel:} The low-mass cut on the massive BH population for different values of $m_{\rm cut}$, for $w=1$ (solid) and $w=2$ (dashed). \emph{Right panel:} The halo mass-BH mass relation: The color coding shows the  BH occupation fraction $p_{\rm occ}$ as a function of the halo mass $M_v$ and the BH mass $m_{\rm BH}$ for a scenario with $f_1 = f_2 = 0.5$, $w_1 = w_2 = 1$, $m_{{\rm cut},1} = 100\Msun$ and $m_{{\rm cut},2} = 10^5\Msun$.}
    \label{fig:pocc}
\end{figure*}

At low redshifts, $z \lesssim 7$, different SMBH seed scenarios are reflected in differences in the low-mass end of the occupation fraction. This is accounted for by the functions $p_j(m_{\rm BH},\vect{\theta}_j)$ for which we consider the following parametric form:
\be \label{eq:pj}
    p_j(m_{\rm BH},\vect{\theta}_j) = \frac{f_j}{2} \left( 1 + {\rm erf}\left[\frac{\log_{10}(m_{\rm BH}/m_{{\rm cut},j})}{w_j}\right] \right)
\ee
with the parameters $\vect{\theta}_j = (f_j,\,m_{{\rm cut},j},\,w_j)$ where $f_j$ is the fraction of SMBHs produced by a given mechanism $j$, $m_{{\rm cut},j}$ characterizes a cut on the minimal SMBH mass, and $w_j$ is the spread with which the minimal-mass cut is applied.\footnote{A qualitatively similar parametrization of the occupation fraction was adopted in~\cite{Miller:2014vta} and in~\cite{Chadayammuri:2022bjj} for studies of SMBH seed models with electromagnetic observations of AGNs.} We fix $\sum_j f_j = 1$. The function $p_j$ is shown in the left panel of Fig.~\ref{fig:pocc} for different values of $m_{\rm cut}$ and $w$. In the following analysis, we consider two cases: (1) a one-component model where the SMBH population arises from one seed population and (2) a two-component model that includes a population arising from light seeds, e.g., BH remnants of Population-III stars, and a population arising from heavy seeds, e.g., BH nuclei of protogalaxies. We show an example of the occupation fraction $p_{\rm occ}$ in the latter case in the right panel of Fig.~\ref{fig:pocc}.~\footnote{There may be other BH binary populations that can complicate the searches of the population related to the assembly of SMBHs. In particular, IMBH binaries may form in dense stellar clusters~\citep{Rasskazov:2019tgb,Fragione:2022avp}. The merger rate of this binary population, however, depends differently on redshift than the merger rate we consider and may become relevant only at $z\lesssim1$.}

We emphasize that the above estimate of the BH merger rate is subject to various uncertainties. In particular, we neglect the additional delays related to the binary evolution following a halo merger~\citep{Hao:2023jfx} as well as possible difficulties for the SMBH to shrink to the galactic cores in dwarf galaxies~\citep{Dayal:2020lri,Dunn:2020pps}, and we extrapolate the halo mass-BH mass relation to much lower masses than where it is currently measured. However, since our goal is mainly to illustrate and compare the capabilities of LISA, AION and AEDGE, a detailed discussion of overall uncertainties is beyond the scope of this work.

\section{GW data analysis}

For each specific SMBH formation model, characterized by the parameters $\vect{\theta}$, we generate $n$ Monte Carlo (MC) realizations of the expected binary populations using Eq.~\eqref{MR}. The number of binaries $N_j$ in any given MC realization follows a Poisson distribution whose mean is the expected number $\bar{N}(\vect{\theta})$ of detectable binaries during one year of observation. The latter is given by
\be
    \bar{N}(\vect{\theta}) =  \int_0^\infty \td \tau \int \td \lambda(\vect{\theta},\vect{x}) \,p_{\rm det}\!\left[{\rm SNR}_c/{\rm SNR}(\vect{x},\tau)\right] \,, 
\ee
where ${\rm SNR}(\vect{x})$ is the signal-to-noise ratio of the GW signal from a binary with parameters $\vect{x}$ in an optimal source-detector system for a given GW detector, $p_{\rm det}$ is the detection probability that accounts for the source location and orientation~\citep{Finn:1992xs,Gerosa:2019dbe}, and ${\rm SNR}_c=8$. 

To characterize how well the parameters of the formation mechanism can be measured, we compute the likelihood for each of the MC realizations~\citep{Mandel:2018mve,Hutsi:2020sol}:
\be \label{eq:likelihood}
    \ell_j(\vect{\theta}) \propto e^{-\bar{N}(\vect{\theta})} \prod_{i=1}^{N_j} \int  \td \lambda(\vect{\theta},\vect{x})\,  P_i (\vect{x}_i|\vect{x})\, , 
\ee
where $\vect{x}_i = (m_{1,i},m_{2,i},z_i)$ are the parameters of the $i$th binary in the generated population, $\vect{x} = (m_{1},m_{2},z)$ are the integration variables and
\be
    \td \lambda(\vect{\theta},\vect{x}) \equiv \frac{1}{1+z} \frac{\td V_c}{\td z} \frac{\td R_{\rm BH}(p_{\rm BH}, \vect{\theta})}{\td m_1 \td m_2} \td m_1 \td m_2\td z \,.
\ee
The distributions $P_i (\vect{x}_i|\vect{x})$ account for the accuracy with which the parameters of the binary $i$ can be measured. We assume that $P_i (\vect{x}_i|\vect{x})$ follows a multivariate Gaussian distribution centered around the true values $\vect{x}_i$, and estimate the covariance matrix from the Fisher information, $\Sigma_{kl}(\vect{x}_i) = \Gamma_{kl}^{-1}(\vect{x}_i)$. Finally, we compute the expected likelihood as
\be
    \ell(\vect{\theta}) = \frac{1}{n}\sum_{j=1}^n \ell_j(\vect{\theta})
\ee
and use that to estimate the expected accuracy at which the model parameters $\vect{\theta}$ can be measured.

\begin{figure*}
    \centering
    \includegraphics[width=\textwidth]{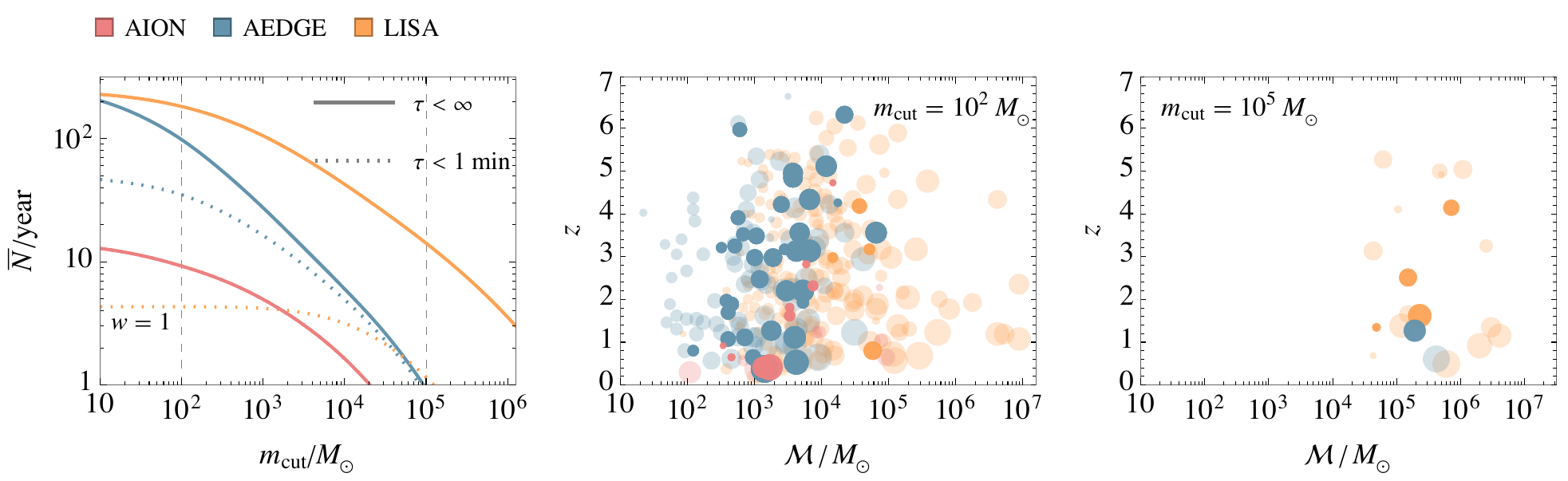}
    \caption{\textit{Left panel:} The expected numbers of binaries detectable by AION-1km, AEDGE and LISA during a year of observation, as functions of $m_{\rm cut}$. The solid curves show all detectable binaries whereas the dotted curves show only those for which the last 1 minute of the merger is seen.
    \textit{Middle and right panels:} Explicit examples of the detectable binaries for a light-seed and a heavy-seed scenario. The sizes of the dots are $\propto \ln[ {\rm SNR}^{-1}]$ with the minimum size corresponding to ${\rm SNR}=10^4$ and the maximal to ${\rm SNR}=10$. The darker dots correspond to binaries for which the last 1 minute of the merger is seen.}
    \label{fig:events}
\end{figure*}

We express the optimal signal-to-noise ratio and the Fisher matrix $\Gamma_{kl}(\vect{x})$ of a signal as~\citep{Poisson:1995ef}
\bea
    &{\rm SNR}(\vect{x}) = \sqrt{\langle \tilde h(\vect{x}) | \tilde h(\vect{x}) \rangle} \,, \\ 
    &\Gamma_{kl}(\vect{x}) = \langle \partial_k \tilde h(\vect{x})|\partial_l \tilde h(\vect{x})\rangle \,,
\eea
by defining the inner product of two complex functions $a(f)$ and $b(f)$ as 
\be
    \langle a(f)| b(f) \rangle = 2 \int_{f(\tau)}^{f(\tau-\mathcal{T})} \!\!\td f \,\frac{a^*(f) b(f) + b^*(f) a(f)}{S_n(f)} \,,
\ee
where $\mathcal{T}$ denotes the observation time and $S_n(f)$ is the noise power spectral density of the considered GW experiment. The partial derivatives in the Fisher matrix are taken with respect to the parameters of the template. We take into account the binary component masses $m_1$ and $m_2$, its redshift $z$, the phase of the GW signals $\phi_c$ and its coalescence time $\tau$, but not the binary sky location or inclination, over which we simply average. To estimate the Fourier transform of the GW strain, $\tilde h(\vect{x})$, we use the inspiral-merger-ringdown template~\citep{Ajith:2007kx}.

\section{Results}

\subsection{Binary population}

We perform the analysis for LISA, AEDGE and AION-1km. As can be seen in the left panel of Fig.~\ref{fig:events}, LISA is expected to observe more binaries than AEDGE. For low-mass seed scenarios this difference is small but it grows with $m_{\rm cut}$ implying that LISA will perform much better than AEDGE for heavy-seed scenarios. In particular, the expected number of binaries observable with AEDGE is less than one for heavy-seed scenarios with $m_{\rm cut} > 10^5 \Msun$.\footnote{We have cut the AEDGE sensitivity at $0.01$\,Hz. Depending on the orbit of the satellites, the Newtonian gravity backgrounds may affect the sensitivity at frequencies below that~\citep{Hogan:2011tsw}. If these backgrounds do not affect the sensitivity, the expected number of binaries for AEDGE will increase significantly, especially in the heavy-seed scenarios. For AION-1km we have considered the low-noise model~\citep{Badurina:2022ngn}.} For Fig.~\ref{fig:events} we have fixed the width of the mass cut to $w=1$, but we have checked that small changes to $w$ do not alter significantly these results. Focusing on binaries that can be observed within 1 minute of the merger, the number of events for LISA is reduced to only $\sim 5 \,{\rm events}$ for the light-seed case. In the heavy-seed case, LISA will not see the last 1 minute of the binaries if $m_{\rm cut} > 10^5\Msun$. On the other hand, AEDGE would detect most of the binaries until within 1 minute of the merger, except for those with $\mathcal{M}_c\lesssim 100 M_{\odot}$. AION-1km has a similar range to AEDGE but, because it has less sensitivity, the number of events is decreased to around $10$ if $m_{\rm cut}\sim10^2M_{\odot}$ and further reduced to 1 event on average for $2\times 10^4\, M_{\odot}$. 

An explicit realization of the binary population in a light-seed scenario with $(m_{\rm cut}=10^2\, M_{\odot})$ is shown in the middle panel Fig.~\ref{fig:events}, and in a heavy-seed scenario with $(m_{\rm cut}=10^5\, M_{\odot})$ in the right panel of Fig.~\ref{fig:events}. We see that AEDGE and, to a lesser extent, AION-1km will probe IMBH binaries over a range of masses that are too light for LISA, covering the mass range covered by terrestrial laser interferometers from $\mathcal{M}_c<100M_{\odot}$ to $\sim 10^4 M_{\odot}$. LISA as expected, will probe heavier binaries up to $\mathcal{M}_c\sim 10^7 M_{\odot}$. The types of events each experiment can detect is quite different. Considering the final stages within one minute of a merger, AEDGE binaries can explore the range $\mathcal{M}\in \left(10^2-10^4\right)\, M_{\odot}$ out to redshifts $z\sim 7$. LISA, on the other hand, observes only the last minutes of heavier binaries with $\mathcal{M}\in\left(10^4-10^6\right)\, M_{\odot}$ and $z\lesssim 4$. Finally, AION-1km will detect a handful of events for $m_{\rm cut}<10^4\, M_{\odot}$, at low redshifts, $z<4$, and mainly in the $\left(10^3-10^4 \right)\, M_{\odot}$ mass range, mostly in the last moments before the merger.

\begin{figure*}
    \centering
    \includegraphics[width=0.84\textwidth]{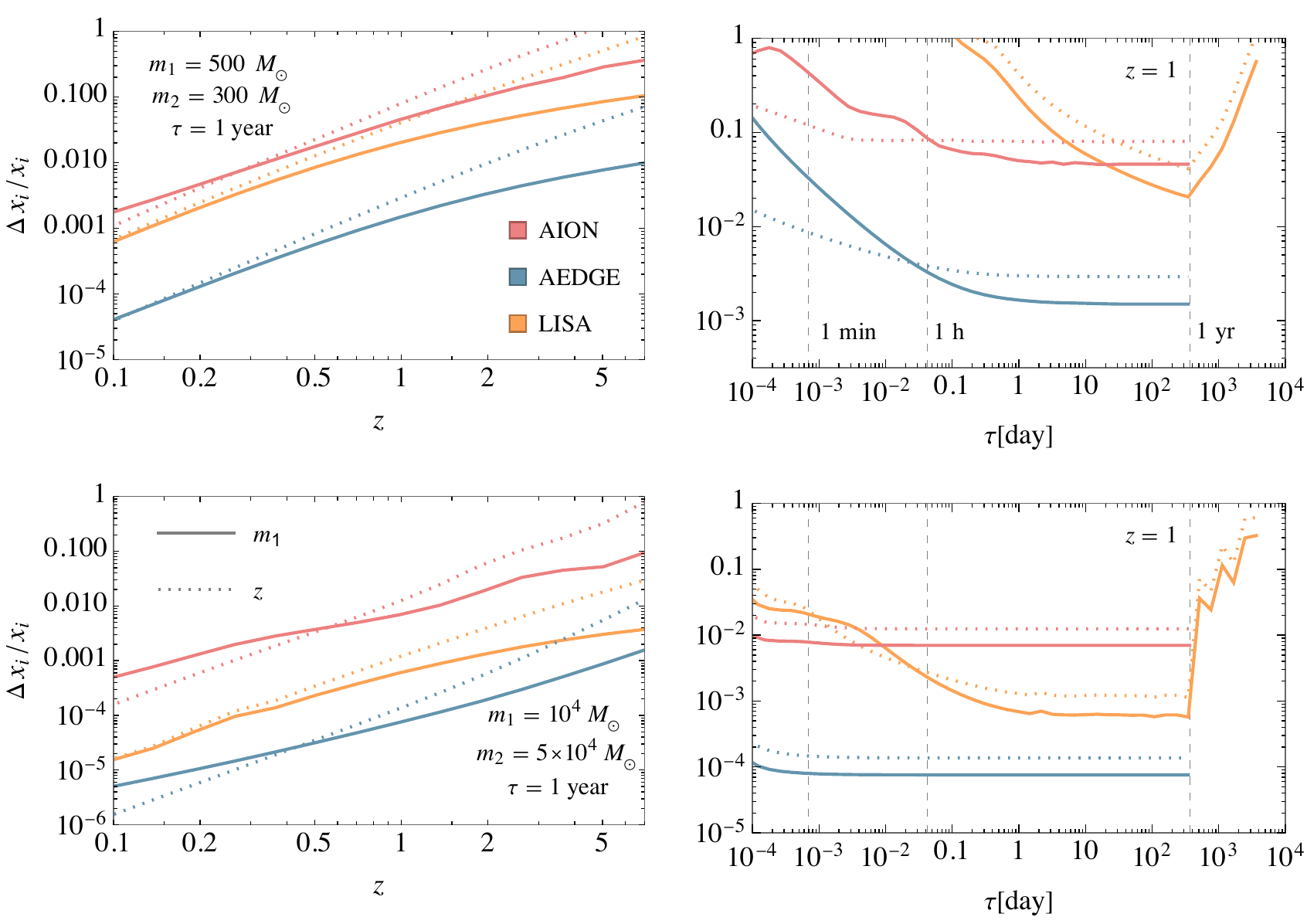}
    \vspace{-3mm}
    \caption{Measurement accuracies for the binary parameters: The upper and lower panels correspond to two binaries whose component masses are fixed. In the left panels, the binary is observed for the last 1 year and the errors are shown as a function of its redshift. In the right panels, $z=1$ and the errors are shown as a function of the binary coalescence time at the beginning of a one-year observation. The errors for the masses of the both BHs are almost the same.}
    \label{fig:measurements}
\end{figure*}

\subsection{Binary parameters}

In Fig.~\ref{fig:measurements} we display the accuracies with which LISA, AEDGE and AION-1km can measure the binary parameters for two representative choices of the binary component masses, a lighter binary with $(m_1,m_2)=(500,300)\, M_{\odot}$ (upper panels) and a heavier binary with $(m_1,m_2)=(10^4,5\times 10^4)\, M_{\odot}$ (lower panels), as functions of the binary redshift $z$ (left panels) and the time to merge  $\tau$ (right panels). In general, the measurement accuracy of the binary parameters $m_1$, $m_2$ and $z$ are similar, and the measurement accuracy is best with AEDGE. The lighter binary enters the AION/AEDGE sensitivity band about a day before the merger. Consequently, the binary is not observable for AION/AEDGE if its coalescence time at the beginning of the experiment is longer than the observation time ($\mathcal{T} = 1\,$yr). LISA instead sees only the inspiralling part until less than a day from the merger and can see binaries whose coalescence time is $\tau \lesssim 10\,{\rm yr}$. The heavier binary is seen with AION/AEDGE for an even shorter time, but since the signal is strong the measurement accuracy is very good. In this case, also LISA sees the merger but the measurement with AEDGE is still more accurate unless the redshift of the binary is high, $z \gtrsim 4$, or its coalescence time is longer than the observation time.

\begin{figure*}
    \centering
    \includegraphics[width=0.9\textwidth]{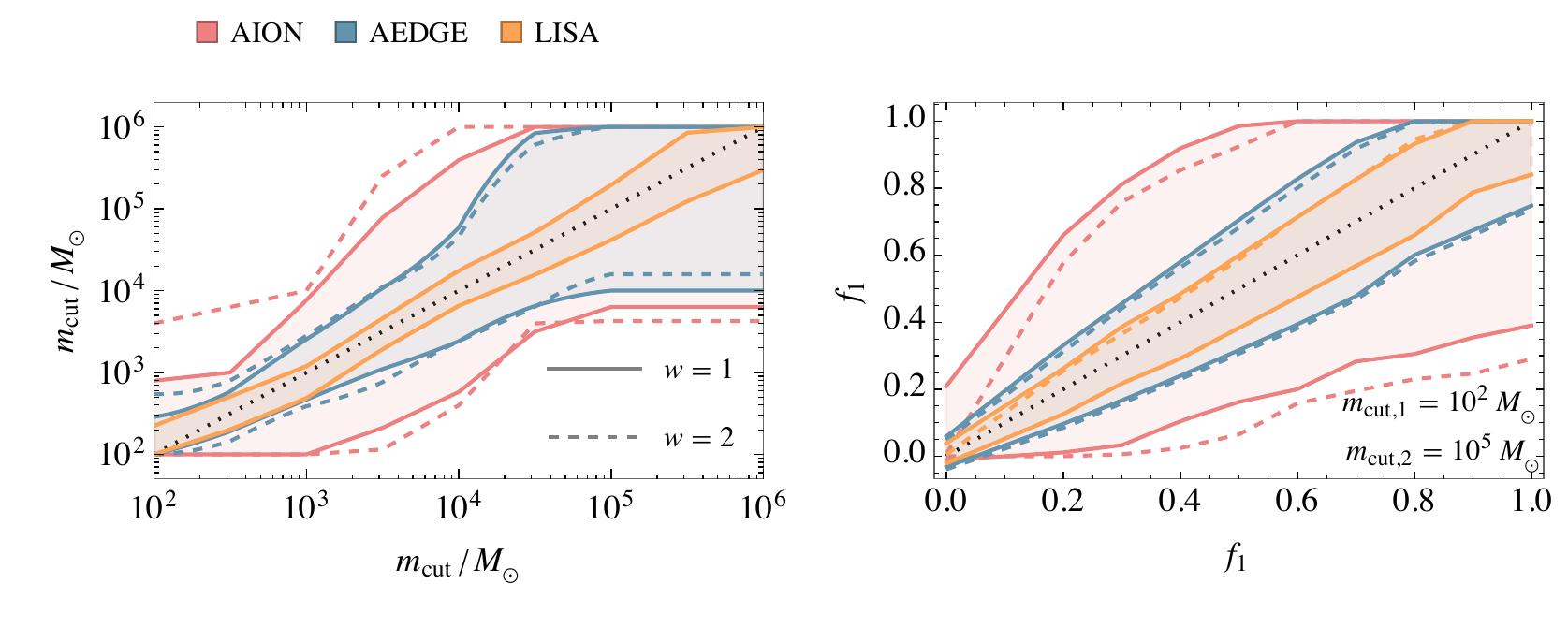}
    \caption{\textit{Left panel:} The 95\% CL accuracy with which LISA, AEDGE and AION-1km could measure $m_{\rm cut}$ over the range $[10^2, 10^6] \Msun$. \textit{Right panel:} The 95\% CL accuracy with which LISA, AEDGE and AION-1km could measure the fraction of light seeds, $f_{\rm 1}$, assuming an input mixture of seeds with masses $10^2$ and $10^5 \Msun$ and $f_2 = 1 - f_1$. The solid and dashed curves in both panels correspond, respectively, to $w=1$ and $w=2$.}
    \label{fig:posteriors}
\end{figure*}

For many binaries, the measurement accuracy with AEDGE or LISA is excellent. The probability distributions of the binary parameters can then be approximated by delta functions in the computation of the likelihood~\eqref{eq:likelihood}, so that the integrals over the binary parameters become trivial because the merger rate is essentially constant over such small parameter ranges. This reduces significantly the computational time for the likelihood. We adopt the delta-function approximation when all binary parameters are measured with better than 10\% accuracy. 

\subsection{Seed scenarios}

To estimate how accurately $m_{\rm cut}$ could be measured with one-year data samples, we have generated $n = 20$ realizations of the detectable binary populations for a given value of $m_{\rm cut}$ considering a single SMBH seed population. The left panel of Fig.~\ref{fig:posteriors} shows on the vertical axis the 95\% CL ranges of $m_{\rm cut}$ estimates for $m_{\rm cut} \in [10^2, 10^6]\, \Msun$ and taking a flat prior for $\log m_{\rm cut}$ over the range $m_{\rm cut} \in [10^2, 10^6]\, \Msun$. The solid and dashed curves correspond to $w=1$ and $w=2$ and show that the measurement accuracy of $m_{\rm cut}$ in these two cases is very similar. We see that the seed mass is best recovered with LISA with a small uncertainty of a factor $\sim 2$. For $m_{\rm cut}\lesssim \times 10^3 \Msun$ the accuracy with AEDGE is similar to LISA, but the 95\% CL range for AEDGE grows for heavier masses. When $m_{\rm cut} \gtrsim 10^5\Msun$ only a lower bound is obtained with AEDGE because the expected number of detectable events is less than one in the heavy-seed scenarios. On the other hand, already AION-1km could measure $m_{\rm cut} \lesssim 3\times 10^4\Msun$ within an order of magnitude at 95\% CL and could place a 95\% CL lower bound of $m_{\rm cut} \gtrsim 4 \times 10^3\Msun$ if $m_{\rm cut} \gtrsim 3\times 10^4\Msun$.

\begin{figure*}
    \centering
    \includegraphics[width=0.4\textwidth]{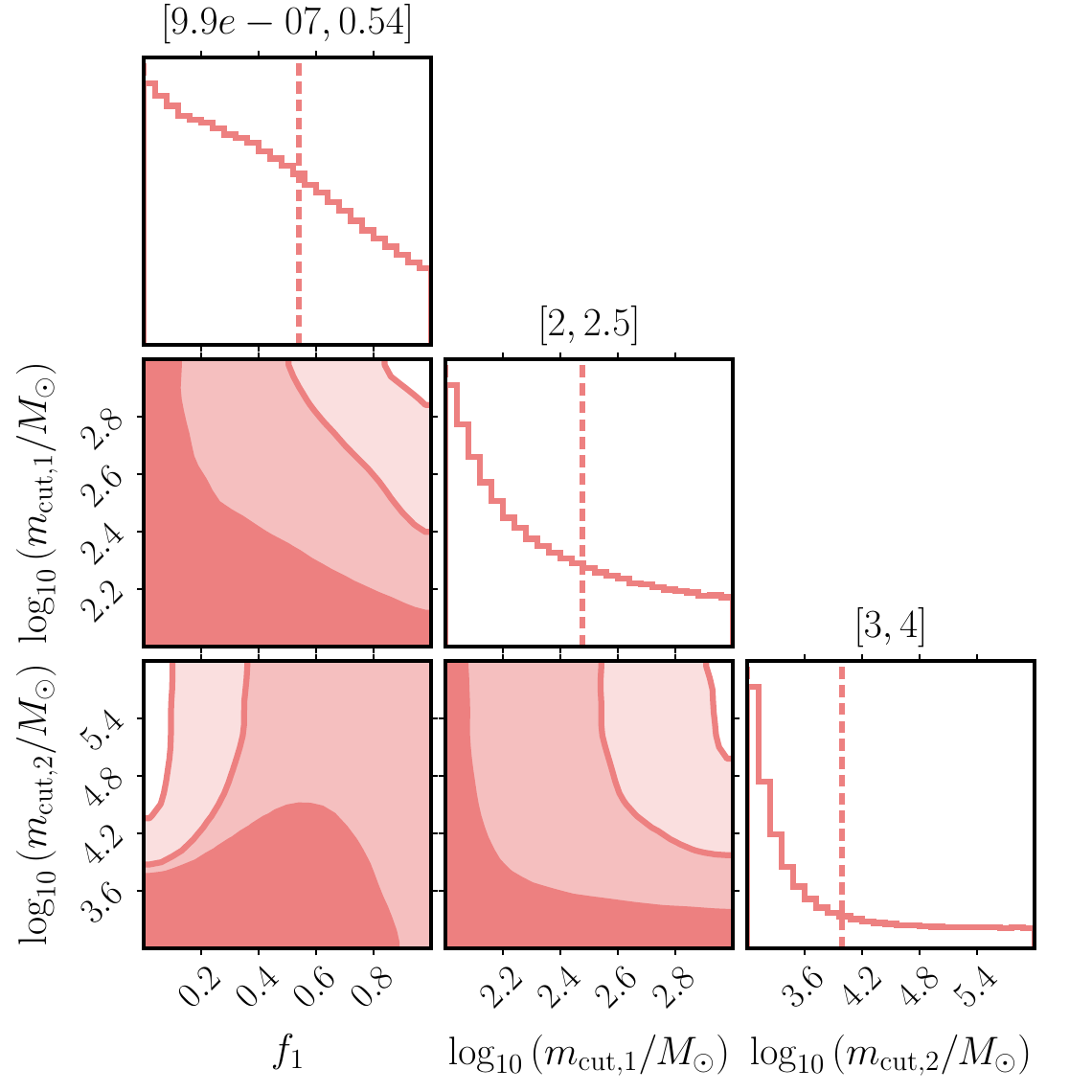}
    \includegraphics[width=0.4\textwidth]{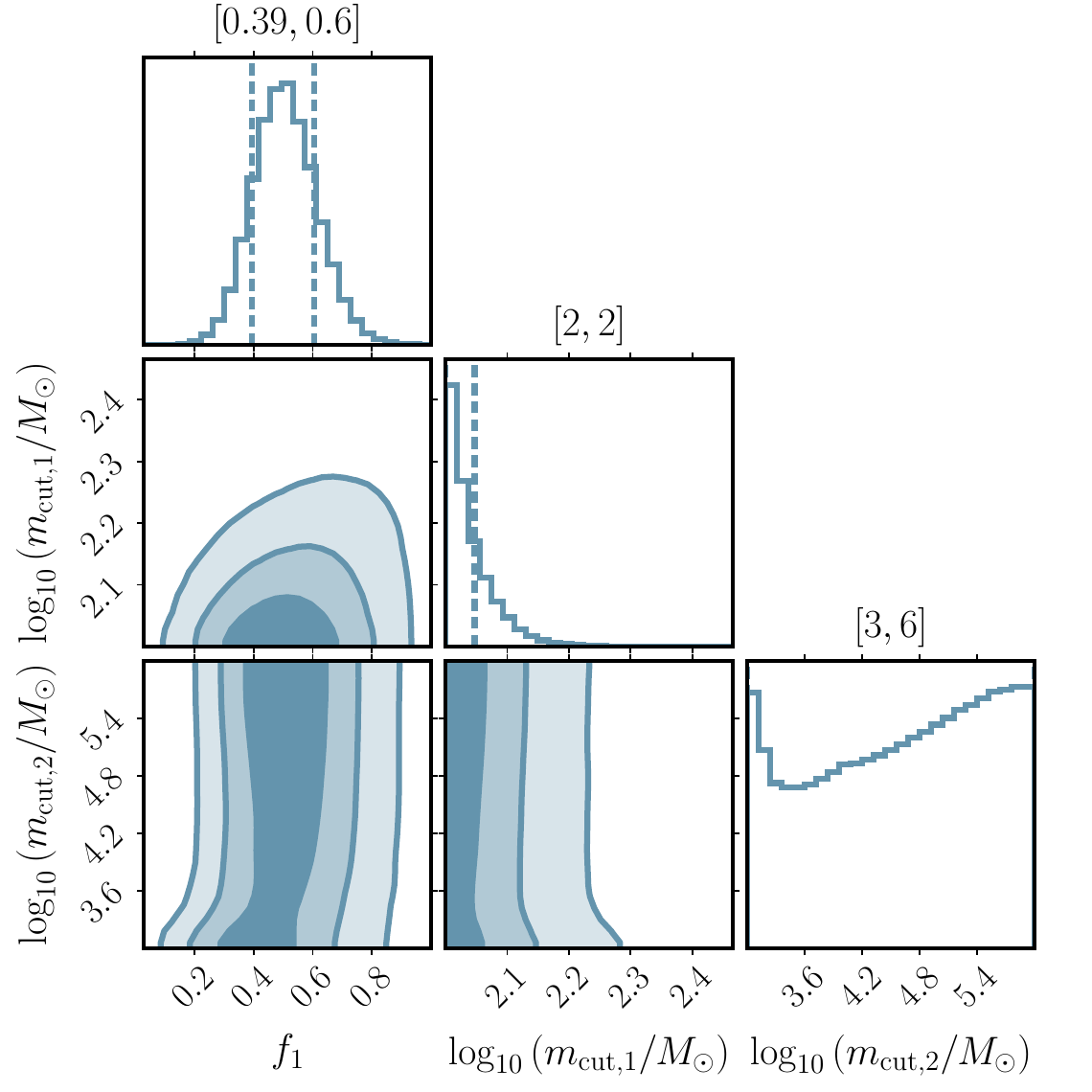}
    \includegraphics[width=0.4\textwidth]{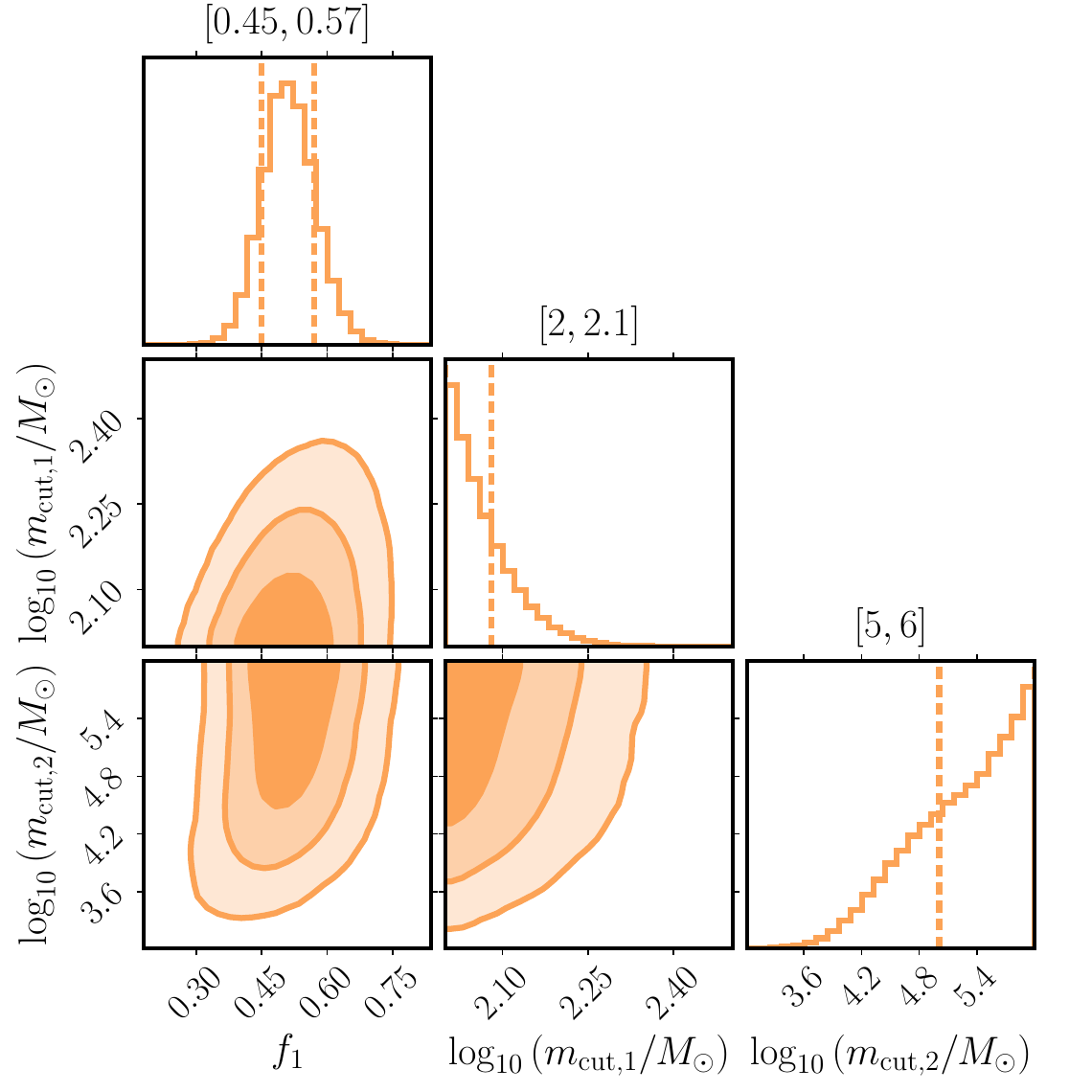}
    \vspace{-2mm}
    \caption{The 1- and 2-dimensional posteriors of the merger rate parameters $\left(f_1, m_{{\rm cut},1}, m_{{\rm cut},2}\right)$ for a two-component seed model with $m_{{\rm cut},1}=10^2\,M_{\odot}$ $m_{{\rm cut},2}=10^5\, M_{\odot}$ and $f_1=0.5$ for AION-1km (top panel), AEDGE (middle panel) and LISA (bottom panel). The contours enclose the 68\%, 95\% and 99\% CL regions and the dashed vertical lines show the 68\% CL ranges in the marginalized posteriors.}
    \label{fig:corner_plots}
\end{figure*}

In the right panel of Fig.~\ref{fig:posteriors}, we have considered the possibility that there are two different SMBH seed populations, a light one with $m_{{\rm cut},1} = 100 \, M_{\odot}$ and a heavy one with $m_{{\rm cut},2} = 10^5 \, M_{\odot}$. We have again generated $n = 20$ realizations of the detectable binary populations for a given value of $f_1$ with $f_2 = 1-f_1$. In the right panel of Fig.~\ref{fig:posteriors} we show how accurately $f_1$ can be recovered at the 95\% CL. We see that LISA could measure $f_1$ with an accuracy $\sim 10\%$, whereas AEDGE could measure $f_{\rm light}$ with an accuracy $\sim 10 - 20 \%$. AION-1km has a much bigger uncertainty but it would be precise enough that it can differentiate at the 95\% CL a pure light-seed from a pure heavy-seed scenario. 

For the right panel of Fig.~\ref{fig:posteriors} we have assumed delta function priors on $m_{{\rm cut},1}$ and $m_{{\rm cut},2}$ that might be theoretically motivated. Instead, in Fig.~\ref{fig:corner_plots} consider a benchmark case with $m_1=10^2\,M_{\odot}$, $m_2=10^5\, M_{\odot}$ and $f_1 = f_2 = 0.5$, and show the one- and two-dimensional posteriors for their measurements with AION-1km (top panel), AEDGE (middle panel) and LISA (bottom panel) assuming the log-uniform priors $2 < \log_{10}(m_{{\rm cut},1}/\Msun) < 3$ and $3 < \log_{10}(m_{{\rm cut},2}/\Msun) < 6$ for the cutoff masses. For simplicity, we keep $w_1 = w_2 = 1$ fixed. We see that AION-1km provides a good estimate of $m_{{\rm cut},1}$, but provides only limited information on $m_{{\rm cut},2}$ and $f_1$. If we would take the prior $m_{{\rm cut},2} \gtrsim 2\times 10^4\Msun$ then we would reproduce the lower bound on $f_1$ with AION-1km seen in the right panel of Fig.~\ref{fig:posteriors}. On the other hand, AEDGE provides good estimates of both $m_{{\rm cut},1}$ and $f_1$ but also does not constrain $m_{{\rm cut},2}$ significantly while LISA not only provides good estimates of both $m_{{\rm cut},1}$ and $f_1$ but also constrains $m_{{\rm cut},2}$ significantly. The marginalized posteriors of $f_1$ for AEDGE and LISA roughly match to the results shown in the right panel of Fig.~\ref{fig:posteriors}.

\section{Conclusions}

We have described in this paper the capabilities of the planned space-borne laser interferometer LISA and the proposed atom interferometers AEDGE and AION-1km to observe mergers of intermediate-mass BHs, measure their parameters, and discriminate between different seed scenarios for the assembly of SMBHs. We have considered the extended Press-Schechter to model the coalescences of galactic halos and estimate a rate for mergers of SMBHs that is compatible with the PTA signals for GWs in the nHz range. We have extrapolated this model to different SMBH seed scenarios by parametrizing the low mass cutoff of the massive BH population. Using this parametrization, we have estimated the possible rates for IMBH mergers, and assessed their detectability and measurability.

We have found that, although LISA has a high rate for observing the early infall stages of IMBH binaries for all the masses studied, this detector loses many binaries as the merger time approaches. Both AEDGE and AION-1km have higher rates than LISA for detections within one minute of the merger. We have shown that AEDGE has the best perspectives for detecting mergers of IMBHs weighing $\lesssim \, 10^4 \Msun$ whereas LISA has better perspectives for IMBHs weighing $\gtrsim \, 10^5 \Msun$. The better detection rates translate into smaller uncertainties in the measurements by AEDGE of binary parameters for IMBHs weighing $\lesssim \, 10^5 \Msun$.

We have estimated the accuracy with which a lower cutoff on the BH seed mass, $m_{\rm cut}$ could be extracted from the prospective GW data. We find that both LISA and AEDGE could determine $m_{\rm cut}$ with precision $\lesssim \, 20 \%$ if $m_{\rm cut} \lesssim \, 10^4 \Msun$, whereas LISA could determine $m_{\rm cut}$ with better precision than AEDGE if $m_{\rm cut} \gtrsim \, 10^4 \Msun$. We also find that both LISA and AEDGE have interesting capabilities for distinguishing between scenarios with different mixtures of seeds with $10^2$ and $10^5 \Msun$. AION-1km could also provide some information, particularly in scenarios with a population of low-mass seeds.

Our results indicate that the space-borne laser interferometer LISA and atom interferometers AEDGE and AION-1km have interesting and complementary capabilities for measuring IMBH mergers and distinguishing between different seed scenarios for the assembly of SMBHs. We should emphasize that our study has been exploratory and should be complemented by an improved modelling of the SMBH seed scenarios and more detailed studies of the instrumental capabilities of GW interferometers. It would also be interesting to extend the analysis to assess the prospects for multimessenger observations and study the prospects for measuring higher-order multipoles of the GW signals that would allow for example for new probes of strong gravity.\\~~\\

\section*{acknowledgments}
The work of J.E. was supported by the United Kingdom STFC Grants ST/X000753/1 and ST/T00679X/1, and that of M.F. was also supported by the United Kingdom STFC Grant ST/X000753/1. The work of J.U. and V.V. was supported by the Estonian Research Council grants PRG803, PSG869, RVTT3 and RVTT7 and the Center of Excellence program TK202. The work of V.V. was also partially supported by the European Union's Horizon Europe research and innovation program under the Marie Sk\l{}odowska-Curie grant agreement No. 101065736.

\bibliography{refs}{}
\bibliographystyle{aasjournal}

\end{document}